\documentclass[11pt]{article}
\usepackage{amsmath}
\usepackage[dvips]{graphicx}
\usepackage{amssymb}
\usepackage{cite}
\newcommand{\be}{\begin{equation}}
\newcommand{\ee}{\end{equation}}
\newcommand{\bear}{\begin{eqnarray}}
\newcommand{\eear}{\end{eqnarray}}

\newcommand{\lapproxeq}{\lower .7ex\hbox{$\;\stackrel{\textstyle  
<}{\sim}\;$}} 
\newcommand{\gapproxeq}{\lower .7ex\hbox{$\;\stackrel{\textstyle  
>}{\sim}\;$}} 
\newcommand{\stackdown}[2]{\lower 1.4ex\hbox{$\;\stackrel{\textstyle{#1}}  
{\scriptstyle{#2}}\;$}}
\newcommand{\beq}{\begin{equation}} 
\newcommand{\eeq}{\end{equation}} 
\newcommand{\refcite}{\cite}

\newcommand{\bea}{\begin{eqnarray}}
\newcommand{\eea}{\end{eqnarray}}

\makeatletter 
\def\slash{\@ifnextchar[{\fmsl@sh}{\fmsl@sh[0mu]}} 
\def\fmsl@sh[#1]#2{%
  \mathchoice 
    {\@fmsl@sh\displaystyle{#1}{#2}}%
    {\@fmsl@sh\textstyle{#1}{#2}}%
    {\@fmsl@sh\scriptstyle{#1}{#2}}%
    {\@fmsl@sh\scriptscriptstyle{#1}{#2}}} 
\def\@fmsl@sh#1#2#3{\m@th\ooalign{$\hfil#1\mkern#2/\hfil$\crcr$#1#3$}} 
\makeatother 
\begin{document}
\begin{titlepage} 
\vspace*{20mm} 
\begin{center} 
\title*{\large{\bf{ Non-critical String Cosmologies}
{\footnote{ 
Talk given at CTP Symposium on Supersymmetry at LHC: 
Theoretical and Experimental Perspectives, Cairo, Egypt 11-14 March, 2007
 } } }}
\\
\vspace{11mm}
 {\bf A.~B.~Lahanas}  \\
\vspace*{6mm} 
  {\it University of Athens, Physics Department,  
Nuclear and Particle Physics Section,   
GR--15771  Athens, Greece\\}
\end{center} 
\vspace*{7mm} 
\begin{abstract}
Non-critical String Cosmologies are offered as an alternative to Standard Big Bang Cosmology. The new features encompassed within the dilaton dependent non-critical terms affect the dynamics of the Universe\'s evolution in an unconventional manner being in agreement with the cosmological data. Non-criticality is responsible for a late transition to acceleration at redshifts z=0.2. The role of the uncoupled rolling dilaton to relic abundance calculations is discussed. The uncoupled rolling dilaton dilutes the neutralino relic densities in supersymmetric theories  by factors of ten, relaxing considerably the severe WMAP Dark Matter constraints, while at the same time leaves almost unaffected the baryon density in agreement with primordial Nucleosynthesis.
\end{abstract}
\end{titlepage} 
\section{\bf{Introduction}}
{{SNIa}} \cite{snIa} and {{WMAP}} 1,3 \cite{wmap} data accumulate in a fast pace confirming 
the existence of {{Dark Energy}} (DE)
which accelerates the Universe and occupies {{ 73 \% }}of its total energy-mass. 
On the other hand, the most recent measurements by WMAP confirmed to unprecentented accuracy the existence 
of {{Dark Matter}} (DM), occupying {{25 \%}} of its total energy-mass, while baryonic density is small
 ( 4 \% ).
But who ordered DE and DM ?  
Fundamental physics theories, notably String Theories,  
are in need to explain these fundamental issues and address to 
the following questions. 
Why  $\Omega_{DE}$ is small ?, This is the  
 {\em{Naturallness problem}}.  
Why $\Omega_{DM} /  \Omega_{DE} \sim {\cal{O}} (1) ? \;$ This is termed as the    
{\em{Cosmic Coincidence problem }}.    

Besides these, the matter-energy density of the  Universe is close to its critical density, $\Omega=1.01 \pm 0.01$, the Hubble constant is accurately known, $h_0=0.73 \pm 0.03$, and various cosmological data have reached a high level of accuracy which must be observed by any cosmological model invoked to explain the evolution of our Universe.  

For the origin of DM 
Supersymmetry (SUSY),  an indispensable part of String Theories, provides good candidates for 
Cold Dark Matter (CDM). WMAP3 data give $\; \Omega_{DM} \; h_0^2 = 0.1045^{+0.0072}_{-0.0095}\; $ 
and SUSY models naturally yield values in the right ball park  $\; \Omega_{CDM} \; h_0^2 \sim 0.1 \; $, for 
a typical supersymmetry scale $M_s \sim \; {\cal O}(TeV)$, the leading candidate being a neutralino 
provided it is the lightest supersymmetric particle (LSP)
\cite{EHNOS}. 

As for the origin of DE there are various proposals. 
A positive {{cosmological constant}}, $\Lambda$, in the Einstein action is the simplest assumption. Its pressure is negative, $p_\Lambda= - \Lambda$, in agreement with the  bounds put on the equation of state of DE, $w_{\Lambda} = -0.97$. 
Another proposal is that a {{quintessence}} scalar field is the carrier of DE \cite{quint}. This must have a small mass $m_q \approx 10^{-33}\;eV$ and it  
does not seem to have an obvious place in any fundamental theory of particle physics.
Also 
the {{dilaton}} occuring in String Theories may be the carrier of DE. Models both in the weak 
( $\phi \longrightarrow -\infty$ ) and non-perturbative limit ( $\phi \longrightarrow +\infty$ )  
\cite{gasperini} of String Theory have been proposed 
( see Ref.~\cite{Gasperini:2002bn} and Refs. therein ). 
The latter surpass the limits put by fifth force experiments yielding naturally small coupling to ordinary matter at late eras.  There are also other proposals like for instance modifications to General Relativity, Braneworld scenarios, Topological defects, and so on which are invoked to expain this fundamental issue. 
Last, but not least, 
The {{rolling dilaton}} in the Q-cosmology scenario \cite{aben,emninfl} offers 
an alternative framework establishing the Supercritical ( or non-critical)  String Cosmology, 
or SSC for short. This opens a new window towards understanding the evolution of our Cosmos which we shall  discuss in the following.

\section{\bf{Supercritical String Cosmology }}

In the feramework of string theories Antoniadis, Bachas, Ellis and Nanopoulos  \cite{aben} 
initiated construction of cosmological string solutions, at the critical 
string dimension, which can be Robertson-Walker-Friedmann (RWF) Geometries in four dimensions ( $4-D$).  
The contributions to central charge is as follows. The 
$4-D$ space-time fields contribute  {{$4-{\delta c}_{RW}$}},     
ghosts yield {$-26\;( \;\mathrm{or} \;-10\; )$ and the     
 "Internal" space fields should contribute $22 \; (\; \mathrm{or} \;6\;)+{\delta c}_{RW}$
so that the total central charge vanishes, ${c_{total}=0}$,
to maintain conformal invariance. In this framework the equations of motion follow from the vanishing of the beta functions 
\begin{equation}
\quad \quad {\tilde \beta}_i  = 0
\end{equation}
and the resulting cosmological backgrounds we shall hereafter call "Critical Q - Cosmologies" with $Q^2$ being a measure of the central charge deficit  ${\delta c}_{RW}$. 

In the framework of the Supercritical String dilaton Cosmology (SSC), Ellis, Mavromatos and Nanopoulos  
\cite{emninfl}  studied more complicated  backgrounds, off the critical dimensions,  that did not satisfy the conformal invariance conditions. They need Liouville dressing to restore conformal invariance. 
Identifying \cite{emn} the zero mode of the Liouville field \cite{ddk} with the target time  results to the modified conditions 
\begin{equation}
\quad \quad {\tilde \beta}_i  = -  {\tilde{\cal{G}}}_i \; .
\end{equation}
The r.h.s of these equations is the result of the non-criticality and play an important role in the equation describing the evolution of the Universe which will be discussed later. The resulting cosmologies we shall term  "non-Critical Q - Cosmologies" to distinguish them from those obtained in the critical case 
studied in Ref.~\cite{aben}. 

In the string frame the bosonic part of the $4 - D$ effective Lagrangian is given by 
\begin{equation}
M^{-2} \;  \sqrt{-G} \; \left[ e^{-2 \phi}  \left(
- R_G + 4 {(\partial_\mu \phi )}^2 - 2  Q^2 \right ) - V(\phi)
\right] + \sqrt{-G} \; {\cal L}_{matter} \; .
\label{lagr}
\end{equation}
The central charge deficit and the charge $Q$ in \ref{lagr} are related by \cite{aben}
$$
\delta c = -3 \;Q^2 \; .
$$ 
In the Lagrangian above we have allowed for matter terms and also for an additional potential term $V(\phi)$ that accomodates the string loop corrections.
The equations of motion are given by
\begin{equation}
\begin{aligned}
&{\tilde \beta}_{\phi} \equiv M^{-2}\;\left( R_G+2 Q^2 +  4 \; {(\partial_\mu \phi )}^2 
- 4 \; \Box \phi - \frac{e^{2 \phi}}{2}\; V^\prime  \right)\; = \;0    \\
&{\tilde \beta}_{\mu\nu} \equiv M^{-2} \left( -R^G_{\mu \nu} + 2 \; \nabla_\mu \nabla_\nu \phi 
+ G_{\mu \nu}  \frac{e^{2 \phi}}{4} \; ( 2 V + V^\prime ) - \frac{e^{2 \phi}}{2 M^2}\; {\tilde T}_{\mu \nu}
\right) \;=\; 0 \; .
\end{aligned}
\end{equation}
In the non-critical string the r.h.s of these equations receive non-vanishing contributions getting the 
form ( For a review see Ref.~\refcite{emnw} ).
\begin{equation}
{\tilde \beta}_i \;=\; - \;M^{-2}\;( g_i^{\prime \prime} + Q g_i^\prime ) \equiv -  {\tilde{\cal{G}}}_i
\label{noncr}
\end{equation}
with $g_i=\phi, G_{\mu \nu}$ and $g_i^{\prime} = d g_i/dt_s$, where $t_s$ is the time in the string frame. 
The r.h.s of \ref{noncr} are the non-critical terms which were absent in the considerations of Ref.~\refcite{aben}. 

In the critical case, ${\tilde{\cal{G}}}_i$ are set to zero and by combining the (00) and (jj) equations,  assuming a flat RWF Universe, one gets  
\begin{equation}
3 \; H^2 \;=\; 8 \pi G_N \rho_m + \rho_\phi \; .
\label{hub}
\end{equation}
where $H$ is the Hubble expansion rate. 
The energy density 
$\rho_m$ includes any sort of matter and  radiation while $\rho_\phi$ refers to the corresponding density for the dilaton. The string frame density and pressure  $\rho_s, p_s$, as read from the energy-momentum tensor ${\tilde T}_{\mu \nu}$, are related to the density, pressures, $\rho_m, p_m $, in the Einstein frame by 
\begin{equation}
\rho_m = e^{4 \phi} \rho_s \quad , \quad p_m=e^{4 \phi} p_s \; .
\end{equation}
The total potential scales by the same factor in this frame given by
\begin{equation}
{\hat V}_{tot} = e^{4 \phi} \; V_{tot} 
\end{equation}
where $V_{tot} =2 Q^2 e^{-2 \phi} + V$. The dilatonic energy density $\rho_\phi$, in the normalization for $\phi$ and the potential the Lagrangian \ref{lagr} is written in, is
$$
\rho_\phi={\dot \phi}^2 + \frac{{\hat V}_{tot}}{2} \; .
$$
Following Ref.~\refcite{diamandis2}
it proves convenient to use a dimensionless Einstein time $t_E$ related to the cosmic time by
$t_E = \omega \; t  $. That done the equation \ref{hub} can be cast in the form 
\begin{equation}
3 \; {\hat H}^2 \;=\; \frac{8 \pi G_N}{\omega^2} \rho_m +
 {(\frac{d \phi}{d t_E})}^2+\frac{{\hat V}_{tot}}{2 \omega^2 }  \; .
 \label{xana}
\end{equation}
In this the dimensionless Hubble expansion rate is related to the actual Hubble rate by ${\hat H}= H/\omega $. 
Furthermore it is convenient to choose $\omega = \sqrt{3} H_0$, so that Eq. \ref{xana} becomes 
\begin{equation}
3 \; {\hat H}^2 \;=\; {\tilde \rho}_m + {\tilde \rho}_\phi \; .
\end{equation}
In these units the densities $ {\tilde \rho}_m, {\tilde \rho}_\phi $ are dimensionless 
\begin{equation}
  {\tilde \rho}_m  \equiv \Omega_m \quad ,  \quad {\tilde \rho}_\phi={(\frac{d \phi}{d t_E})}^2+\frac{{\hat V}_{tot}^0}{2 } \; .
\end{equation}
$ {\tilde \rho}_m $ is actually the ratio of the density $ {\rho}_m $ to the critical density, often denoted by $\Omega_m $.
The potential $V_{tot}^0 \equiv 2 {\hat Q}^2 e^{-2 \phi} + V^0 $ with $V^0 \equiv V/ \omega^2$ and ${\hat Q} \equiv Q/\omega$ are also dimensionless. In this system of units the cosmological equations can be cast as a system of first order differential equations which can be treated numerically,  
\begin{equation}
\begin{aligned}
&2 {\dot {\hat H}} \;+ \; {\tilde \rho}_m +{\tilde p}_m +{\tilde \rho}_\phi +{\tilde p}_\phi \;=\;0   \\
&\ddot \phi \; + \; 3 \;{\dot \phi} \;{\hat H} +\frac{    {{\hat V}_{tot}^{0^\prime} }     }{4} + 
\frac{1}{2} \;( \;{\tilde \rho}_m - 3 {\tilde p}_m \;)\;=\;0 \\
&{\dot {\tilde \rho}}_m \;+\; 3 \; {\hat H} \; ( \;{\tilde \rho}_m +
{\tilde p}_m \;) - {\dot \phi} \;( \;{\tilde \rho}_m - 3{\tilde p}_m \;)\;=\;0 \; \; .
\end{aligned}
\end{equation}
Given the equation of state for each species involved in ${\tilde \rho}_m $
these are solved with initial values for ${\dot \phi}^0, {\hat H}^0,{{\tilde \rho}_m}^0 $. The superscripts denote their values today. One can trade the deceleration $q^0$, whose value is experimentally known,  for ${\dot \phi}^0$ using the relation
\begin{equation}
q = -1 + \frac{1}{{\hat H}^2} \; ( \; {\dot \phi}^2 + 
\frac{1}{2}\;( \;{\tilde \rho}_m +{\tilde p}_m \;)\;)
\end{equation}
A few remarks are in order. The first is that the initial value
of the dilaton can be taken zero if the string loop  corrections to the potential are neglected, i.e.  $V^0(\phi)=0$, which we assume in the following. The system is then invariant under a dilaton shift followed by a  Q-charge rescaling
$$
\phi \rightarrow \phi + c \quad, \quad {\hat Q} \rightarrow e^{-c}\;{\hat Q} \; .
$$ 
Then any solution with $\phi^0 \neq 0$ is mapped to another with vanishing 
initial value for the dilaton $\phi^0 = 0$ and a rescaled charge ${\hat Q}$. The second remark 
concerns the equation $3 \; {\hat H}^2 \;=\; {\tilde \rho}_m + {\tilde \rho}_\phi$ which yields 
\begin{equation}
e^{2 \phi}\;{\hat Q}^2 + {\dot \phi}^2 + {\tilde \rho}_m +\frac{{\hat V}^0}{2} -3 {\hat H}^2 =0 \; .
\label{qqq}
\end{equation}
One can verify that the constancy of ${\hat Q}$ follows from this equation by taking the derivative of both sides. This equation can be also used to express the initial value  
${\hat Q}_0$ in terms of the remaining inputs.
Since we have taken ${\hat H}_0 =1/\sqrt{3} $, in these system of units,  it follows that  
{\em{only the densities and the deceleration $q_0$ are needed to solve the set cosmological equations !}}

From the positivity of ${\hat Q}^2_0$ and ${\dot \phi}^2_0$ bounds on today's value of the deceleration $q_0$ are obtained. In fact one has 
$$
-1 + \frac{3 \Omega_M}{2} + 2 \Omega_r < q_0< 2 - \frac{3 \Omega_M}{2} -  \Omega_r \; .
$$
For values of matter density close to those observed, $\Omega_{M}=0.24$, this yields $\; q_0 > -0.65 \;$ marginally allowing deceleration values in the range $q_0 \approx -0.60 $ as observed experimentally. With $\Omega_{M}=0.24$, $q_0 \approx -0.60$, and zero cosmological constant, it follows 
that ${\dot \phi}_0^2 \approx 0.015$ and ${\hat Q}^2_0 \approx 0.755$
resulting to 
$$
w_\phi =
{\left( \frac{p_\phi}{\rho_\phi} \right)}_{today} \approx -1 \; .
$$
Therefore in this scheme the DE is carried by the dilaton field and today's dilaton energy is mainly potential whose value is controlled by a nonvanishing central charge. 
A problem that arises in this model is that 
the deceleration $q(z)$, as function of the redshift, hardly agrees 
with current observations that acceleration started at 
$z \approx 0.15 - 0.2$. This is remedied in the non-critical case which we shall disuss in the following.

In the SSC the modified Friedmann equations, in the same system of units, 
are~\cite{diamandis2,Lahanas:2006xv}
\begin{equation}
\begin{aligned}
&3 \; {\hat H}^2 - {\tilde{\rho}}_m - {\tilde \rho}_{\phi}\;=\; \frac{e^{2 \phi}}{2} \; \tilde{\cal{G}}_{\phi}   \\
&2\;\dot{\hat H}+{\tilde{\rho}}_m +{\tilde  \rho}_{\phi}+
{\tilde{p}}_m +{\tilde p}_{\phi}\;=\; \frac{\tilde{\cal{G}}_{ii}}{a^2}   \\
&\ddot{\phi}+3 {\hat H }\dot{\phi}+ \frac{{{\hat V}_{tot}^{0^\prime} }}{4} 
+ \frac{1}{2} \;( {\tilde{\rho}}_m - 3 {\tilde{p}}_m )= 
- \frac{3}{2}\; \frac{ \;\tilde{\cal{G}}_{ii}}{ \;a^2}- \,
\frac{e^{2 \phi}
}{2} \; \tilde{\cal{G}}_{\phi}  
\end{aligned}
\end{equation}
The terms on the r.h.s. are the non-critical terms which for lack of space we do not display explicitly. 
In Ref.~\refcite{diamandis} consistent cosmological solutions of these equations, in the absence of matter, were sought which tend asymptotically, in cosmic time, to the conformal backgrounds considered in Ref.~\refcite{aben}. 
In Refs.~\refcite{diamandis2}, \refcite{emmn} it was shown that off-equilibrium supercritical string cosmologies,
are consistent with the current astrophysical data.

In the non-critical case the charge  
$\hat Q$ varies with time and its variation is provided by the Curci-Paffuti 
$\sigma$-model renormalizability constraint. 
In terms of the non-critical terms ${\tilde{\cal{G}}}_{\phi,ii}$ this is given by 
\begin{equation}
\frac{d \tilde{\cal{G}}_{\phi} }{d t_E} \;=\; - 6\; e^{\;-2 \phi}\;( {\hat H}+ \dot{\phi} ) \;
\frac{ \;\tilde{\cal{G}}_{ii}}{ \;a^2} \; \; .
\end{equation}
Combinining the available equations we get the continuity equation as in the critical case
\begin{equation}
\frac{d {\tilde{\varrho}}_m }{dt_E}+ 3{\hat  H} ( {\tilde{\varrho}}_m +{\tilde p}_m) 
- \dot{\phi}\;({\tilde{\varrho}}_m - 3 {\tilde{p}}_m )\;=\;
6\;({\hat H}+\dot{\phi})\; \frac{ \;\tilde{\cal{G}}_{ii}}{a^2} 
- 2 \;{\hat Q } \dot{\hat Q} e^{2 \;\phi}  \; \; .
\end{equation}
We then split the total density ${\tilde{\varrho}}_m $ to matter ("dust") $\varrho_b$,  radiation, 
$\varrho_r$, and the rest, $ \varrho_e $, which we coin ("exotic"), 
having equations of state with parameters $w_b=0, w_r=1/3$ and $w_e=\mathrm{undetermined}$. The latter is assumed constant, for simplicitly, but other more involved options are certainly available. Then we get three separate continuity equations for 
$\varrho_b, \varrho_r, \varrho_e $ given by 
\begin{equation}
\begin{aligned}
&\frac{d {{\varrho}}_r }{dt_E}+ 4 {\hat H}  {{\varrho}}_r  \;=\;0  \\
&\frac{d {{\varrho}}_b }{dt_E}+ 3 {\hat H}  {{\varrho}}_b - \dot{\phi} {{\varrho}}_b  \;=\;0  \\
&\frac{d {\varrho_e}}{dt_E}+ 3 {\hat H} \;(1 + w_e)\;{\varrho}_e  
- \dot{\phi}\;( 1 - 3\;w_e )\;{\varrho}_e\;=\; 
&6\;({\hat H}+\dot{\phi})\; \frac{ \;\tilde{\cal{G}}_{ii}}{a^2} 
- 2 \;{\hat Q} \dot{\hat Q} e^{2 \;\phi} \; .
\label{con3}
\end{aligned}
\end{equation}
From the first of these it becomes apparent that 
radiation does not explicitly feel the presence of the dilaton unlike dust which does feel it through a dissipative term $- \dot{\phi} {{\varrho}}_b  $, as is seen from the second of Eqs. \ref{con3}. The "exotic" matter $  {{\varrho}}_e  $ feels both the effect of the dilaton and the non-critical terms. In deriving these we have tacitly assumed that the non-critical terms affect only the exotic piece but neither dust nor radiation. 
Ignoring string loop corrections to potential, $V^0(\phi)=0$, we get a first order 
system having as dependent variables 
{{${\dot \phi}, {\hat H},{e^\phi {\hat Q}}$}} and the densities 
 {{ $ \rho_{b,r,e}\; $}}, 
\begin{equation}
\begin{aligned}
&\ddot{\phi}= -2 \hat{H}^2-3 \hat{H} \dot{\phi} - e^{\phi} {\hat Q}( \dot{\phi}+\hat{H})+
\;\frac{1}{2} \rho_b +\frac{2}{3} \rho_r + \frac{(1+w_e)}{2} \rho_e \; \nonumber  \\
&3 \dot{\hat{H}}=-\hat{H}^2-2 {\dot{\phi}}^2+e^{\phi} {\hat Q} ( \dot{\phi}+\hat{H})
-  \;\frac{3}{2} \rho_b - \frac{5}{3} \rho_r - \frac{(3+w_e)}{2} \rho_e \;  \\
&{ \dot{{\varrho}}_e }+
2 {\hat Q} \dot{{\hat Q}} e^{2 \phi} =  
 -3 \;(1+w_e)\; \hat{H}  {{\varrho}}_e 
+ ( 1 - 3 w_e) \; \dot{\phi}\;{{\varrho}}_e  \nonumber \\
& + 4 \;(\hat{H}+\dot{\phi})\;
( -\hat{H}^2+ {\dot \phi}^2 + e^{\phi} {\hat Q}( \dot{\phi}+\hat{H}) + w_e \rho_e + \frac{\rho_r}{3}\; )\nonumber  \\
&\dot{\rho}_b \;=\;  -3 \hat{H} {\rho}_b + \dot{\phi}\;{\rho}_b  \nonumber \\
&\dot{\rho_r}\;=\;-4 \hat{H} {\rho}_r   \nonumber \\
&{( e^\phi \;{{\hat Q}})}^{\cdot} \;=\; F(e^\phi {\hat Q},{\hat H},{\dot \phi},\rho_i)   \nonumber \\
\end{aligned}
\end{equation}
The r.h.s of the last equation we do not write it explicitly for lack of space. 
This system is invariant under {{$\phi \rightarrow \phi + c\;$}}, 
{{$ \hat Q \rightarrow e^{-c} \; \hat Q \;$ }}.  

In order to solve them and obtain acceptable backgrounds we have to specify the initial conditions. 
These are
\begin{itemize}
\item{Today's Hubble constant ${\hat H}_0$, which in the units we are working it has the value ${\hat H}_0 =1/\sqrt{3}$, and the value of the deceleration today  $q_0$. }
\item{The values of the densities ${ \rho}_b, {\rho}_r, {\rho}_e$
 today.}
\item{The dilaton's value $\phi_0 $ today which we take vanishing. As stated previously 
a non-vanishing dilaton value would merely rescale the charge $Q$ due to the symmetry 
$\phi \rightarrow \phi + c,  \hat Q \rightarrow e^{-c} \; \hat Q \;$   
.}
\end{itemize}
The derivative of the dilaton field ${\dot{\phi}}_0$, which is also needeed, is fixed by these 
inputs as in the critical case through its relation to the deceleration. Also the central charge deficit ${\hat Q}_0$ is determined  due to an algebraic relation  
( similar to that obtained in the critical case ),  which follows from Friedmann's 
equations,  given by 
$$
2 \;{\hat Q}^2 - e^{- \phi}{\hat  H} \;{\hat Q} + e^{- 2 \phi}\; 
( \;{\dot{\phi}}^2 - 8 {\hat H}^2 - 3 {\hat H} \dot{\phi}+ 
\frac{5}{2} { \rho}_b + \frac{8}{3}{ \rho}_r \;+ \frac{5+w_e}{2} { \rho}_e ) \;=\;0  \; .
$$
Today's value of the exotic matter energy-density we take vanishing, 
but other options, less attractive, are available. The  value of  $w_e$  is undetermined 
and it is actually a fitting parameter in this approach.

\begin{figure}[t]
\begin{center}
\includegraphics[width=8.0cm]{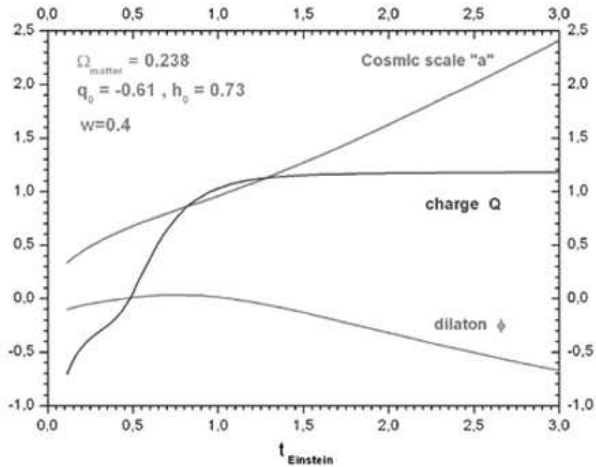}
\vspace*{8pt}
\end{center}
\caption{
The cosmic scale factor $a$, the central charge deficit charge $\hat{Q}$, and the dilaton $\phi$, as functions of the Einstein time $t_E$ for the inputs shown in the figure.
} 
\label{f1a}
\end{figure}
\begin{figure}[t]
\begin{center}
\includegraphics[width=8.0cm]{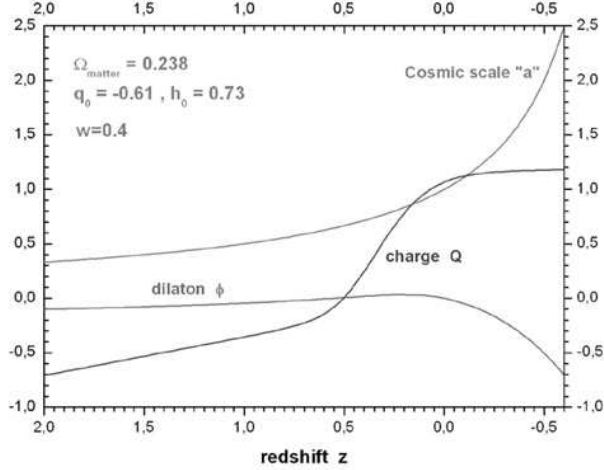}
\vspace*{8pt}
\end{center}
\caption{
The same quantities as in figure \ref{f1a} plotted as functions of the redshift $z$ for $-0.5 < z < 2.0$.
} 
\label{f1b}
\end{figure}
\begin{figure}[t]
\begin{center}
\includegraphics[width=8.0cm]{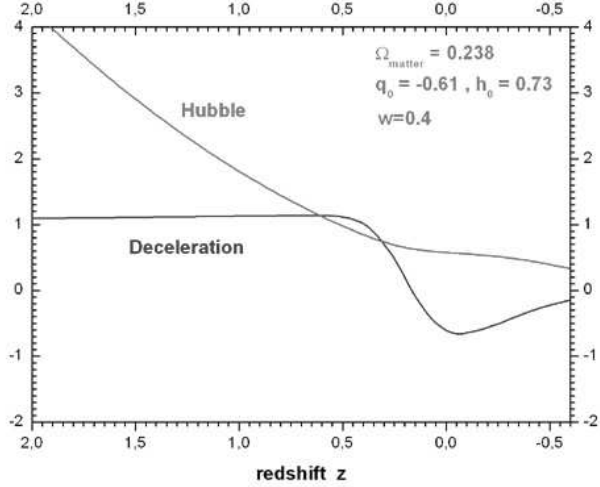}
\vspace*{8pt}
\end{center}
\caption{
The deceleration $q$ and the dimensionless Hubble expansion rate, ${\hat H} = H / {\sqrt{3} H_0}$, as functions of the redshift $z$.
} 
\label{f2}
\end{figure}
\begin{figure}[t]
\begin{center}
\includegraphics[width=8.7cm]{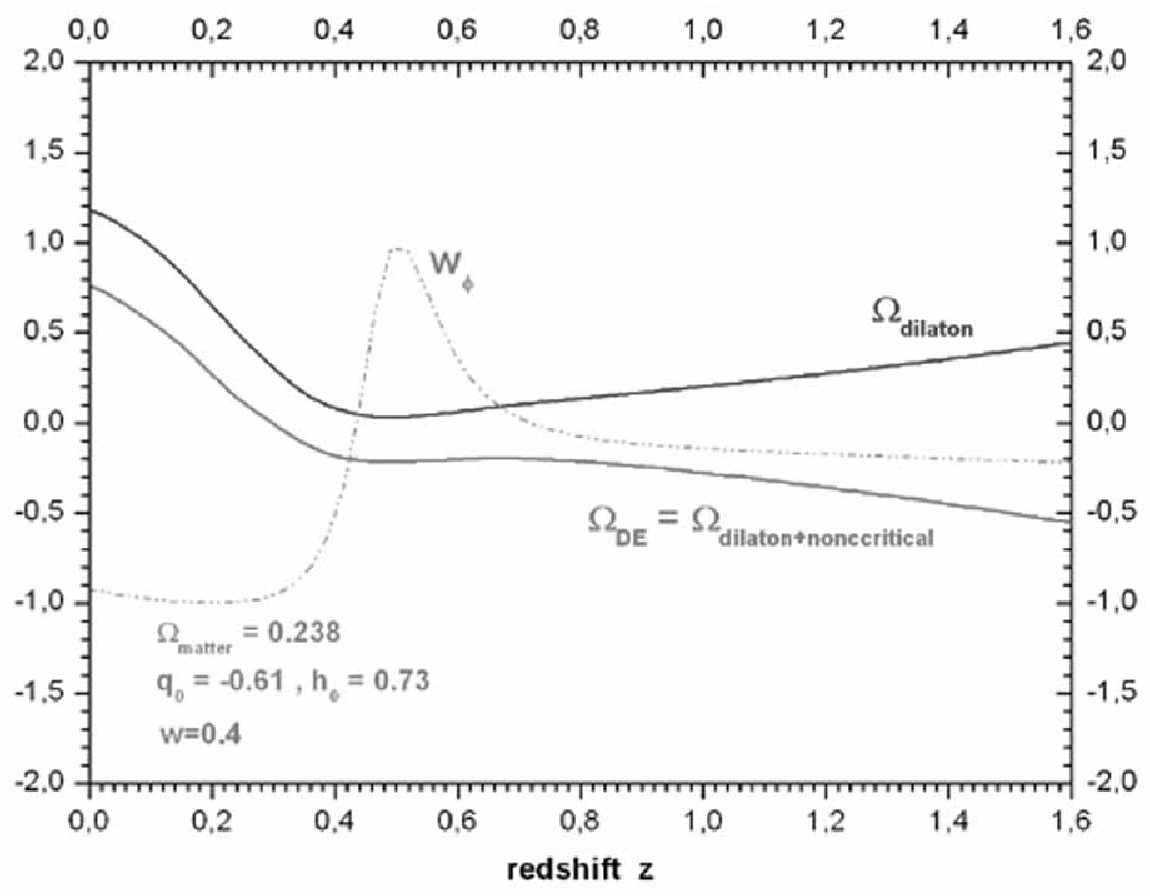}
\vspace*{8pt}
\end{center}
\caption{
The vacuum energy, $\Omega_{DE}\equiv \Omega_{dilaton}+\Omega_{noncritical}$, the dilaton energy,  
$\Omega_{dilaton}$, and the equation of state function $w_{\phi}$ as functions of the redshift.
} 
\label{f3}
\end{figure}
In Fig. \ref{f1a} we display the solutions obtained for the dilaton $\phi$ the cosmic scale factor $a$ and the charge $Q$ for a value $w_e=0.4$, as functions of the Einstein time $t_E$. The deceleration has been taken equal to $q_0=-0.61$ and the total matter density $\Omega_{matter}=0.238$. The density of the exotic matter today and the dilaton's value are taken vanishing as we have already discussed.  For other values of $w_e$ we obtain different solutions. However the particular value is in agreement with a smooth variation of the vacuum energy and Nucleosynthesis as we shall discuss later on. Today corresponds to a value for $t_E$ equal to $t_E \approx 1.07$ and the solutions obtained approach their asymptotic values for values of $t_E$ larger than 1.4. The same quantities as functions of the redshift$z$, for $z=-0.5 - 2.0$, are shown in 
Fig. \ref{f1b}. 
The deceleration $q$ and the Hubble expansion rate, in units of $\sqrt{3} H_0$, are shown in Fig. \ref{f2}.
 Note that the entrance to acceleration phase occurs at redshifts around $z \approx 0.2 $ in good agreement with the current astrophysicsl observations. In figure \ref{f3} we display the dilaton's energy, in units of the critical density, $\Omega_{dilaton}$, and the Dark Energy, $\Omega_{DE}$, interpreted as the energy carried by the dilaton and the non-critical terms, for the inputs shown in the figure. In this figure we also display the value of the function $w_{\phi}$ of the equation of state for the dilaton. The latter approaches values close to -1, for redshifts smaller than $0.2$, indicating that the dilaton's energy is mainly potential energy in this regime. It should be remarked that the Dark Energy is smooth for redshift values $z=0.0 - 1.6$ in egreement with the recent observations of supernovea \cite{riessnew}

\section{\bf{Relic abundances and the rolling dilaton}}

The continuity equation for any species $i$, matter or radiation, in the presence of the dilaton is 
$$
\dot{\rho}_i
 +3 \hat{H} (\;{\rho}_i + p_i ) -{\dot \phi}\; ({\rho}_i - 3 p_i )=0 \; . \nonumber 
$$
For temperatures 
$T >> m_i$ \; matter is relativistic and $\; \; p_i \approx \rho_i / 3 $. Therefore  $\dot \phi$ decouples from the continuity equation which takes on the form of the first of Eqs. \ref{con3}. On the other hand for 
$T < m_i \;$ matter is non-relativistic, $p_i \approx  0$, and 
$\dot \phi$ contributes through a dissipative term  $- {\dot \phi}\; \rho_i$. In this case we get the second of Eqs. \ref{con3}.
For massive particles of mass $m_i$, as long as we are 
in the temperature regime $m_i > T > T_{today} $, the energy density can be written as  $\rho_i = n m_i$ and 
in order to account for the presence of the dilatonic dissipative term,  the Boltzmann equation should  be modified accordingly \cite{diamandis2,lmnbolt}:
\begin{equation}
\frac{dn}{dt} + 3 H n + < v \sigma > (n^2 -n^2_{eq}) - {\dot \phi}\;n \;=\;0 \; .
\end{equation}
If $T \;$ is the photon gas temperature, as  measured by antennas and satellites, and $\rho_r \; $ the radiation density including all relativistic particles at a given epoch then \cite{kolb}
$$
\rho_r \;=\; \frac{\pi^2}{30} \; g_{eff}(T) \; T^4
$$
with $ g_{eff}$ counting the relativistic degrees of freedom. On the other hand the redshift $z$ 
is related to the temperature through the following relation 
$$
z+1 \;=\; {\left( \frac{g_{eff}}{g_{eff}^0 }\right)}^{1/4} \; \frac{T}{T_0} \; .
$$
Today's values, labelled by $0$, are 
$\;g_{eff}^0 = 2 + (7 N_{\nu}/ 4) ( {T_{\nu}}/{T_0} ) \approx 2.91$ 
and $ T_0 \approx 2.7 ^0K\; $. 
In the minimal supersymmetric model (MSSM)  and for temperatures larger than the typical supersymmetry breaking scale, $T > M_{SUSY}$, we have  
$$
z+1 \; \approx \; 1.27 \times 10^{13}  \; \frac{T}{GeV} \; .
$$
Therefore if decoupling of a SUSY cold dark matter candidate, say a neutralino $\tilde{\chi}$, occurs at 
$T_{\tilde \chi} \approx m_{\tilde \chi}/20$  we need probe redshift regions as large as $z \approx 10^{15}$ to know the effect of the $\dot \phi$ term. 
It proves convenient to define ${\tilde g}_{eff}$ through the relation 
$$
\rho_{tot} \;=\; \frac{\pi^2}{30} \; {\tilde g}_{eff}(T) \; T^4 \; .
$$
In this 
$\rho_{tot}$ is the total contribution to the energy-matter density, appearing on the r.h.s  of the equation 
$3 { H}^2 = 8 \pi G_N \rho_{tot}$, which also includes the dilaton and the non-critical terms. 
Then, as we have shown in Ref.~\refcite{lmnbolt}, the quantity $\dot{\phi}/H$ and the ratio $Y$  defined by  
\begin{equation}
Y \equiv \frac{{\tilde{g}}_{eff}}{ g_{eff}} = 
\frac{1}{H_0^2} \; \frac{ {\hat{H}}^2}{\Omega_r} \nonumber
\end{equation}
control the modifications to freeze-out temperature and relic density. 
These quantities are calculated after solving the cosmological equations.
For a particle species  $i$ of mass $m_{i}$, the freeze out point 
$x_f = T_f / m_i$ is found to be \cite{lmnbolt} 
\begin{equation}
\begin{aligned}
&x_f^{-1}\;=\; \\ 
&ln \left[ 0.03824 \; g_s\; \frac{M_P m_i}{\sqrt{g_{eff}^{*}}} \;
x_f^{1/2} {<{ v \sigma}> }_f 
\right]\;-\;
\frac{1}{2} \;  ln  \; Y(x_f) \;+\;
\int_{x_f}^{x_{in}} \;  \frac{{\dot \phi} { H}^{-1}}{x} \;dx 
\label{xf}
\end{aligned}
\end{equation}
In this equation the superscript $*$ denotes values at $x_f$ and  $x_{in} \approx {\cal{O}} (1)$ is the point above which the particle is relativistic and $\dot \phi$ does not contribute. Results are insensitive to its precise value because of the smallness of $1/H$ for high redshift values.  
The additional terms appearing in Eq. \ref{xf} affect the conventional $x_f$ calculations by only $10 \%$ for SUSY or hadron species. 
However for the relic density things are more dramatic since it is found that  \cite{lmnbolt} 
\begin{equation}
\Omega h_0^2 \;=\;R \; \times \left( \Omega h_0^2 \right)_{(0)} \; 
\label{redu}
\end{equation}
with the prefactor $R$ given by 
\begin{equation}
R\;=\; 
exp \; 
\left[ \; \int_{x_0}^{x_f} 
\frac{{\dot \phi} {{H}}^{-1}}{x} \; dx 
+ \frac{1}{2}\; ln \; Y(x_f)
\right] \; .
\label{rrr}
\end{equation}
\noindent 
${( \Omega_{} h_0^2 )}_{(0)}$ is the relic density derived in ordinary treatments, usually approximated by
\begin{equation}
\left( \Omega_{} h_0^2 \right)_{(0)}\;=\;\
\frac{1.066\; \times  10^9\; {GeV}^{-1}  }{ M_P \;\sqrt{g^{*}_{eff}} \;J} 
\end{equation}
\begin{table}[]
\begin{center}
{\begin{tabular}{c|cccc} \hline 
   && LSP && Hadron \\
  \hline \hline \\
$\frac{1}{2}\; ln \;Y(x_f) $ && 1.84  &&2.51 \\
 \\
$ \int_{x_0}^{x_f} \frac{{\dot \phi} {H}^{-1}}{x} \; dx $ &&  -4.12  && -3.14 \\
 \\
R \hphantom{0} &&  0.102 &&  0.53  \\ \hline
\end{tabular} }
\end{center}
\caption{
{The values of the quantities controlling the factor R of Eq. \ref{rrr} for 
 $w_e=0.4$. 
The cases for an LSP and a typical hadron are shown along with the resulting values for $R$}
}
\label{ta1}
\end{table}
where $J\equiv \int_0^{x_f} <{ v \sigma}> dx$.  Typical outputs of the quantities $\frac{1}{2}\;ln \;Y(x_f)$ 
and $\int_{x_0}^{x_f} \frac{{\dot \phi} {H}^{-1}}{x} \; dx$ controllling the factor $R$ in Eq. (\ref{redu}), 
for values of the parameter of the equation of state $w_e = 0.4 $, are given in Table \ref{ta1} for a   neutralino, which is assumed to be the LSP, and a typical hadron. These quantities combine to yield a factor $R$ which for the case of the LSP is of the order of $0.1$ while for a typical hadron is of order one. 
The results for a hadron are independent of SUSY inputs, since the freeze out temperature is much less than the SUSY breaking scale, $ T_H \approx \; 10^{-2} \; GeV << M_{SUSY}$. 
The results for the LSP have a mild dependence on SUSY inputs since in this case the freeze out temperature is smaller, but not much smaller, than  $M_{SUSY}$, $ T_{LSP} \approx m_{LSP}/20  \; < M_{SUSY}$.

The factor $R$ depends on $w_e$, however we'd better use  values around $w_e \approx 0.4$ which are cosmologically preferred. We have found that for such values radiation prevails over matter  
at $T \approx 1\; MeV$, as demanded by Primordial Nucleosynthesis and also that Dark Energy $\Omega_{DE}$, as the energy carried by the dilaton and the non-critical terms, has a rather smooth $z$- dependence in the range $0 <z < 1.6$ in agreement with recent cosmological data. 

\begin{figure}[t]
\begin{center}
\includegraphics[width=8.7cm]{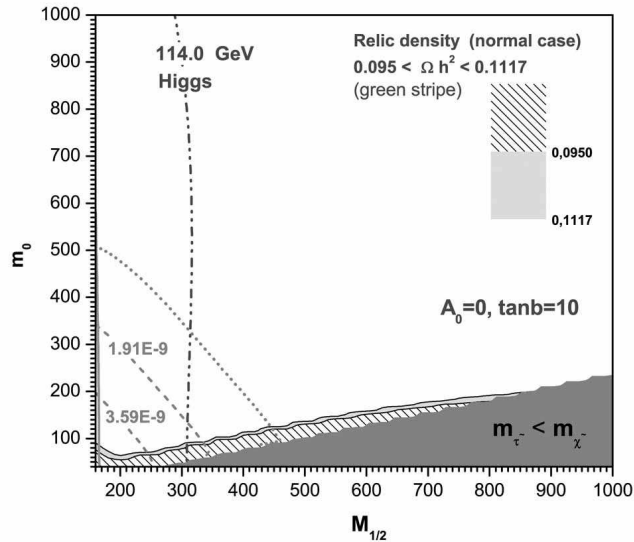}
\vspace*{8pt}
\end{center}
\caption{
In the thin grey stripe the neutralino  relic density is within the  WMAP3 limits
$0.0950 < \Omega_{CDM}h^2 < 0.1117$, for values $A_0=0$ and $tan \beta=10$, according to the conventional calculation. The dashed double dotted line delineates the boundary along which the Higgs mass is equal to $114.0 \; GeV$. The dashed lines are the $1\sigma$ boundaries for the allowed region put by the $g-2$ muon's data. The dotted lines are the same boundaries at the $2 \sigma$'s level. In the hatched region $0.0950 > \Omega_{CDM}h^2$, while in the dark region at the bottom the LSP is a stau.}
 \label{f4} 
\end{figure}
\begin{figure}
\begin{center}
\includegraphics[width=8.7cm]{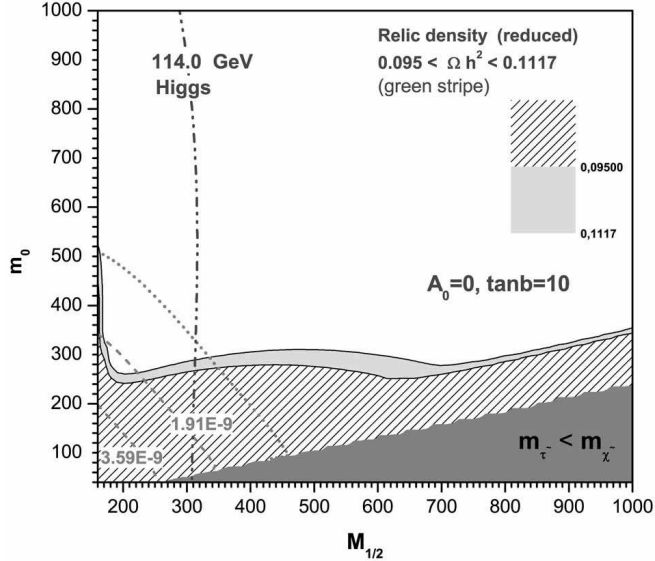}
\vspace*{8pt}
\end{center}
\caption{
The same as in figure \ref{f4} according to the new calculation in which the relic density is reduced
as described in the main text.}
\label{f5}
\end{figure}
The WMAP3 data for Dark Matter impose severe constraints on some of the supersymmetric models  
\cite{susyconstr} and the  dilution of the neutralino Dark Matter by ${\cal {O}} (10)$ may have  
dramatic phenomenological consequences. Supersymmetric models have been extensively studied in the literarure 
\cite{bbik,coannihil,Baer:2002fv,tril,funnels,higgspole,focusit,Baer:2006ff}. 
As the relic density is reduced to its $10 \%$ value, as compared to that obtained in ordinary treatments, the cosmologically allowed region is expanded and shifted to regions of the  $m_0, M_{1/2}$ plane that is forbidden in the conventional approaches \cite{Lahanas:2006xv}.
In Fig. \ref{f4} we display, in the $m_0, M_{1/2}$ plane, the cosmologically allowed region as a thin stripe above the hatched area, for the SUSY inputs shown in the figure. The constrained supersymmetric scenario is assumed where all scalar masses are common at the unification scale. Also common are the gaugino masses and the trilinear couplings. 
\begin{figure}[t]
\begin{center}
\includegraphics[width=8.7cm]{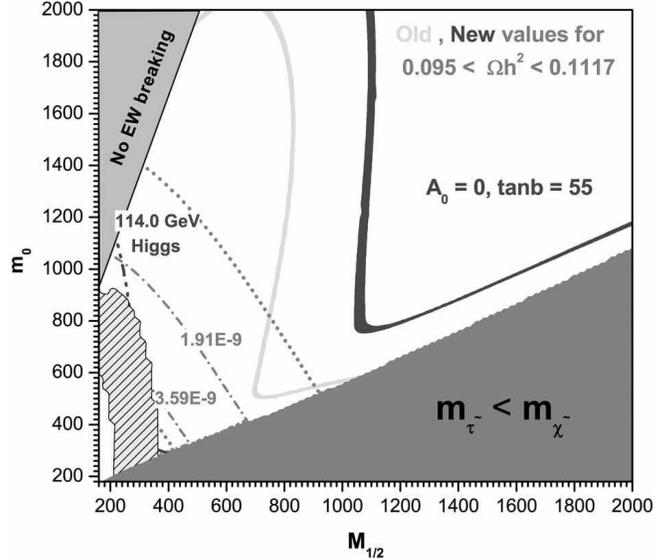}
\vspace*{8pt}
\end{center}
\caption{
In the very thin light grey stripe the neutralino  relic density is within the  WMAP3 limits
$0.0950 < \Omega_{CDM}h^2 < 0.1117$, for values of $A_0=0$ and $tan \beta=55$, according to the conventional calculation. The thin darker V-shaped stripe, lying to the right of it, is the same region according to the new calculation. The MSSM inputs are shown in the figure. The Higgs mass  and $g-2$ boundaries are as in figure \ref{f4}. The hatched shaded region on the left is excluded by $b \rightarrow s \; \gamma$ data.
} 
\label{f7}
\end{figure}
In the figure the lower mass Higgs bound, put by LEP last measurements, and the bounds imposed by the muon's anomalous magnetic moment measurements are also shown~\cite{eidel}. In Fig. \ref{f5} the same figure is shown but with the modified neutralino relic density accoding to Eq. \ref{redu}. One sees a shift of the cosmologically allowed region upwards which relaxes the severe constraints of the conventional models,  allowing for a broader region of the parameter space.  
In Fig \ref{f7} we changed the input value for the angle $\tan \beta=10$ to $\tan \beta=55$. 
The thin light grey stripe, almost vertical to the grey shaded area excluded by $m_{\tilde {\tau}} < m_{\tilde \chi}$ constraint, is the cosmologically allowed region according to the conventional calculation. The darker and broader V-shaped stripe to the right of it, which extends to higher values of $M_{1/2}$, is the area allowed according to the new calculation. On the left of the figure the hatched area designates the area excluded by the  $b \rightarrow s + \gamma$ constraint. 

\section{\bf{Conclusions}}
Non-critical or Supercritical String Cosmologies (SSC) provide an alternative viable framework to  
describe the evolution of our Universe. Within this framework 
\begin{itemize}
\item{
The derived cosmological equations are in agreement with the cosmological data 
and Dark Energy is carried by the dilaton and the non-critical terms.
}
\item{
The transition to accelerating phase occurs naturally at redshifts $z \sim 0.2 $, in agreement with astrophysical observations. 
}
\item{
Matter density is affected by the rolling dilaton through a dissipative pressure term $\sim {\dot \phi}$ diluting significantly relic abundances of supersymmetric CDM candidates while it leaves unaffected the predictions for ordinary matter.
}
\item{
Dark Energy evolves rather smoothly for the last ten 
billion years while it fits  the available astrophysical data including Primordial Nucleosynthesis. 
}
\end{itemize}
The effect of the non-critical terms to the 
Cosmic Microwave Background anisotropies, inflation and other relevant 
issues need be further studied. 
Variants of the simplest scenario can be considered: 
Non-trivial dilaton charges for matter and/or radiation, non-constant equation of state 
for the exotic matter and inclusion of loop corrections of the effective String Theory to the dilaton potential, may further improve the early-time predictions as we approach the Big Bang singularity. 

\section*{Acknowledgments}
This work is supported by funds made available by the European Social Fund (75\%)
and National (Greek) Resources (25\%) - EPEAEK~B - PYTHAGORAS. \\
The material of this lecture is based on work done in collaboration with  B.C.~Georgalas, G.A.~Diamandis, N.E.~Mavromatos and  D.V~Nanopoulos, 

\end{document}